\documentclass[10pt,a4paper]{iopart}
\usepackage{amssymb}
\usepackage{wrapfig}
\usepackage{graphicx}
\usepackage[usenames]{color}

\begin{document}
\title[F\"orster resonant energy transfer between molecules and Rydberg atoms]{State resolved investigation of F\"orster resonant energy transfer in collisions between polar molecules and Rydberg atoms}
\author{F. Jarisch$^1$}
\author{M. Zeppenfeld}
\ead{martin.zeppenfeld@mpq.mpg.de}
\address{Max-Planck-Institut f\"ur Quantenoptik, Hans-Kopfermann-Stra{\ss}e~1, 85748 Garching, Germany}
\address{$^1$Present address: Fraunhofer AISEC, Parking 4, 85748 Garching, Germany}


\begin{abstract}
We perform a comprehensive investigation of F\"orster resonant energy transfer in a room-temperature thermal mixture of ammonia molecules and rubidium Rydberg atoms. Fully state-resolved measurement of the Rydberg-atom populations is achieved by combining millimeter-wave state transfer with state-selective field ionization. This allows aspects of the energy transfer process such as state dependence, ammonia pressure dependence, and dependence on the energy resonance condition to be investigated in detail. Our results pave the way for future quantum experiments combining polar molecules and Rydberg atoms.
\end{abstract}



\maketitle

\section{Introduction}
Quantum hybrid systems are attracting considerable attention since the combination of two different quantum systems ideally profits from the advantages of both systems while allowing the disadvantages to be mitigated by the complementary system. Thus, for example, the combination of atomic and molecular ions using quantum-logic spectroscopy allows the high level of control achievable for atoms to be transferred to molecular ions~\cite{Wolf16,Chou17}. This provides access to the rich internal structure of molecular ions for various applications.

For neutral but polar molecules, a combination comparable to molecular and atomic ions is the combination with Rydberg atoms. Here, the large electric dipole moments in both systems facilitate interactions over macroscopic distances, making this combination favorable for various applications. For example, interactions with Rydberg atoms have been suggested to allow cooling of both the motional as well as the internal degrees of freedom of molecules~\cite{Zhao12,Huber12}. Rydberg atoms might be used for reading out the internal molecular states and for nondestructive molecule detection~\cite{Kuznetsova16,Zeppenfeld17}. A hybrid molecule-Rydberg-atom system would be a versatile platform for quantum information processing~\cite{Kuznetsova11}. Finally, bound states between Rydberg atoms and molecules should allow the investigation of long-range polyatomic Rydberg molecules~\cite{Rittenhouse10,Aguilera17}. However, to date practically no experimental work in this direction has been performed.

A particularly powerful mechanism for coupling polar molecules and Rydberg atoms is F\"orster resonant energy transfer~\cite{Foerster48}. F\"orster resonant energy transfer occurs for equally spaced, dipole coupled energy levels in two systems in close spatial proximity. In this case, the resonant dipole-dipole coupling causes an excitation in one system to be transfered to the other system and vice versa. F\"orster resonant energy transfer was first observed in 1922 in the dissociation of hydrogen by collisions with excited mercury~\cite{Cario22}, and is widely used in various areas~\cite{Andrews99}.

Using F\"orster resonant energy transfer to couple molecules and Rydberg atoms has several advantages. First, since the coupling relies on dynamic rather than static dipole moments, no dc electric fields are needed to orient the molecular or atomic dipole moments, typically resulting in larger dipole moments and a correspondingly stronger coupling. Second, dynamic dipole coupling selects on energy separation between the energy levels in both systems, allowing an easier discrimination from background processes. Third, F\"orster resonant energy transfer causes a state change in both systems which persists even after the systems stop interacting. This is much easier to detect than the transient shift of energy levels caused by static dipole-dipole interactions.

F\"orster resonant energy transfer between molecules and Rydberg atoms has been investigated in several instances in the past. In the late 1970's to the early 1990's, this was investigated in the context of studying interactions between polar molecules and Rydberg atoms more generally~\cite{Matsuzawa83,Dunning83}. This includes observation of resonant energy exchange with a number of different molecules species, including NH$_3$~\cite{Smith78,Kellert80a,Petitjean86a,Ling93b}, CH$_4$ and CD$_4$~\cite{Gallagher80}, HF~\cite{Higgs81a,Kalamarides87}, and HCl~\cite{Stebbings81}. In particular, Petitjean {\it et al.} were first to experimentally investigate energy exchange with the inversion mode in ammonia~\cite{Petitjean86a,Petitjean86b}, as is performed here. In very recent work, Zhelyazkova and Hogan were able to observe effects of electric fields on the interaction cross sections~\cite{Zhelyazkova17a,Zhelyazkova17b}.

In this paper, we investigate F\"orster resonant energy transfer between thermal ensembles at room temperature of rubidium Rydberg atoms and ammonia molecules. Combining state-selective field ionization (SFI) with mm-wave state transfer allows to perform an in-depth characterization of the Rydberg atom populations, in particular, the state distrubutions following interactions with molecules. This allows us to measure high quality data illuminating various aspects of the energy transfer process.

Our paper is structured as follows. The experimental setup and related topics are described in section~\ref{section II}. In particular, the mm-wave state transfer is discussed in section~\ref{section IIa}. The main results of this paper are presented in section~\ref{section III}, focusing on the following six aspects of F\"orster resonant energy transfer. First, we show that F\"orster resonance energy transfer results in robust signals with large signal to noise ratio which indicates the presence of ammonia at densities of about $10^{10}\,$cm$^{-3}$. Second, we measure populations in all S, P, and D Rydberg states in the vicinity of an initially excited P state and show that within the experimental error, state transfer due to ammonia occurs only as expected due to F\"orster resonant energy transfer. Third, we investigate populations in individual Rydberg M-sublevels and show that $M$-sublevels populated by F\"orster resonant energy transfer are consistent with dipole selection rules, as expected.  Fourth, we investigate the ammonia-pressure dependence of the F\"orster resonant energy transfer. Fifth, we study the effect of a DC electric field on F\"orster resonant energy transfer and show that changing the electric field can substantially alter the energy transfer rate. Sixth, we show that the energy resonance condition of F\"orster resonant energy transfer can be used to measure a low-resolution spectrum of the ammonia inversion splitting. In section~\ref{section IV}, we consider future directions of research and possible applications of this work.

\section{Experimental Setup}\label{section II}

\begin{figure}[t]
\centering
\includegraphics[width=0.9\textwidth]{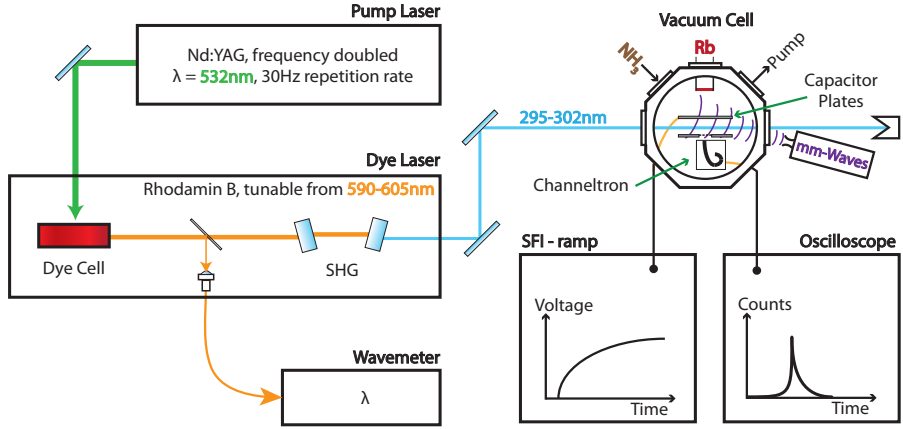}
\caption{Experimental setup, as described in the main text.}\label{fig setup}
\end{figure}

The experimental setup is shown in Fig.~\ref{fig setup}. Rubidium atoms produced in a vacuum cell by applying current to a rubidium dispenser are excited to Rydberg states via one-photon excitation with a frequency-doubled pulsed dye laser (Fine Adjustment, model Pulsare). Typical laser pulse energies are up to $1$\,mJ at the experiment. The laser is collimated (beam waist $\sim1\,$mm) between a pair of capacitor plates ($75$\,mm by $50$\,mm, separated by $5$\,mm), allowing application of homogeneous electric fields to Stark-shift the Rydberg energy levels as well as application of fast high voltage ramps for state-selective field ionization (SFI)~\cite{Gallagher77} of the Rydberg atoms. Fast high voltage ramps are generated by rapidly switching a voltage source with a high-voltage switch (Behlke, model GHTS 30) in combination with an R-C circuit. A hole in the bottom capacitor plate covered by a metal mesh allows electrons produced by SFI to be extracted and detected with a channeltron. The channeltron signal is recorded with a digital oscilloscope (LeCroy, model HRO 66Zi).

Ammonia molecules (NH$_3$) from a gas bottle are injected into the vacuum cell through a needle valve. The vacuum cell is continuously evacuated with a turbo pump, so a constant flow of ammonia is needed to maintain a given ammonia pressure inside the cell, as monitored by a pressure gauge (Pfeiffer, model PKR 261). Background pressure after pumping overnight with no ammonia applied is roughly $10^{-7}$\,mbar. This value slightly increases when applying current to the Rubidium dispenser, and the resulting background pressure is subtracted from pressure readings with ammonia to obtain the partial pressure of ammonia.

\begin{figure}[t]
\centering
\includegraphics[width=0.85\textwidth]{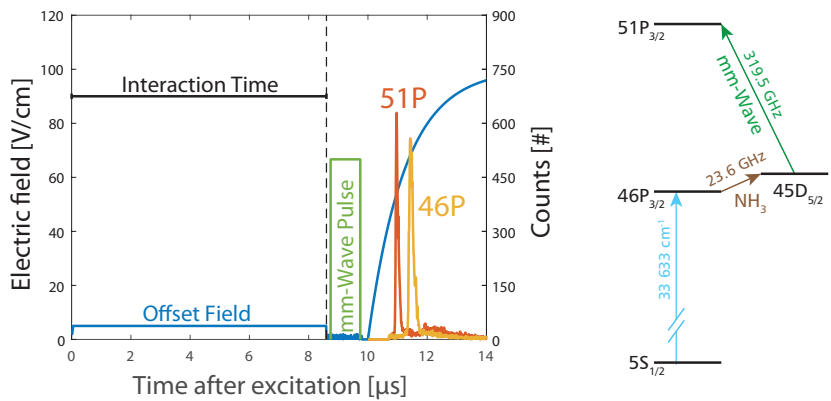}
\caption{Typical experimental sequence and level scheme for rubidium. We show the applied electric field strength versus time as well as the time window for application of the mm-wave pulse. Morover, typical SFI signals with integrated electron clicks versus time are shown for Rubidium excited to the 46P and 51P states, demonstrating that these states are clearly distinguishable based on electron arrival time.}\label{fig sequence}
\end{figure}

The experimental sequence and rudidium level scheme is shown in Fig.~\ref{fig sequence}. Rubidium is initially excited to an $n$P$_{3/2}$ state by the pulsed laser, with $n=46$ for most of the results presented. The $46$P$_{3/2}$ state is chosen because it is separated by $23.6$\,GHz from the $45$D$_{5/2}$ state, roughly corresponding to the inversion splitting of ammonia. As a result, ammonia molecules in the excited state of the inversion doublet can transfer their energy to a Rydberg atom during a collision via F\"orster resonant energy transfer, which should result in population transfer to the state $45$D$_{5/2}$. To allow this to happen, the initial Rydberg excitation is followed by a variable interaction time. During this time, an offset electric field can optionally be applied to the capacitor plates, thereby Stark shifting the Rydberg energy levels. For the results presented in sections~\ref{section III.5} and \ref{section III.6}, this offset field is distinct from the field applied during the laser pulse, allowing the effect of a field during laser excitation to be separated from the effect on collisions with ammonia.

The interaction time is followed by a one microsecond mm-wave pulse. This mm-wave pulse couples two Rydberg levels, for example the $45$D$_{5/2}$ state and the $51$P$_{3/2}$ state in Fig.~\ref{fig sequence}. Since we expect almost no Rydberg population to be in the state $51$P, the mm-wave pulse transfers roughly half of the population in the state $45$D$_{5/2}$ to the state $51$P$_{3/2}$. More specifically, due to decoherences and sufficiently high power, the mm-wave pulse ideally equalizes the population among all coupled M-sublevels (see discussion below). The mm-wave pulse thereby probes the population in a single Rydberg state. This is an essential diagnostic tool and is discussed in detail in the following subsection. After the mm-wave pulse, a high voltage ramp field ionizes the Rydberg atoms, and the resulting electrons are detected with a channeltron.

\subsection{State Analysis via mm-Wave State Transfer}\label{section IIa}

While SFI results in clearly distinct ionization peaks for Rydberg states with sufficiently different principal quantum number $n$, as shown in Fig.~\ref{fig sequence}, this is much less the case for closely lying Rydberg states or for different $M$-sublevels of a given quantum state. The $n$P and $n-1$D states which are to be distinguished to observe F\"orster resonant energy transfer show particularly strong overlap. Obtaining the desired information about the Rydberg state populations directly from the SFI signals would thus be quite challenging.

\begin{figure}[t]
\centering
\includegraphics[width=1\textwidth]{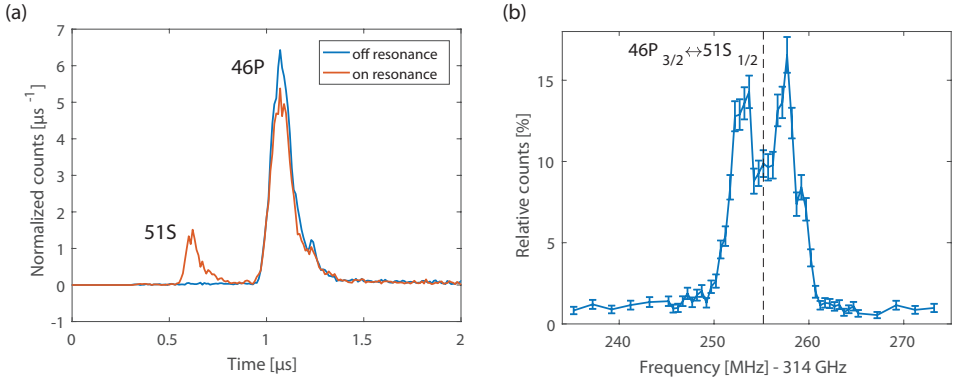}
\caption{Characterization of mm-wave state transfer. (a). SFI signal for laser excitation to the $46$P state, followed by a mm-wave pulse on- or off-resonant to the $46$P$_{3/2}\leftrightarrow51$S$_{1/2}$ transition. The signal is normalized (integrating over time gives unity) to eliminate the effect of variations in laser power. (b). Relative counts in the vicinity of the $51$S peak in (a) as a function of mm-wave frequency. The dashed vertical line indicates the frequency for the $46$P$_{3/2}\leftrightarrow51$S$_{1/2}$ transition expected from quantum defect theory~\cite{Li03,Bhatti81,Marinescu94}. The double-peak structure is due to Zeeman splitting in magnetic fields ($\sim200\,\mu$T) caused by the pressure gauge in the experiment. Low mm-wave power is used for this measurement to avoid saturation broadening.}\label{fig mmwave}
\end{figure}

An extremely useful diagnostic tool to obtain more accurate information about the Rydberg states is the use of a mm-wave source. This allows to selectively drive transitions between Rydberg states with a considerable difference in principal quantum number, allowing individual Rydberg states to be selectively probed as briefly discussed above. For this purpose, an amplifier multiplier chain (Virginia Diodes, model WR10 AMC) connected to a microwave synthesizer roughly covering the frequency range 220-320GHz is placed outside the vacuum cell, with a horn antenna directing mm-wave radiation through a vacuum viewport toward the capacitor plates, as shown in Fig.~\ref{fig setup}.

Using mm-wave state transfer provides two major advantages. First, since the mm-wave radiation substantially changes the Rydberg state energy, the resulting SFI ionization peaks are well isolated, allowing the atoms in the final state to be measured with very low background signal from Rydberg atoms in the initially excited state. Second, since the mm-wave radiation can be switched on/off while leaving all other aspects of the experiment unchanged, and due to the narrow spectral bandwidth of the mm-wave transitions, any change in signal caused by switching the mm-wave radiation is practically guarenteed to result from the Rydberg states being addressed.

Typical data for mm-wave state transfer is shown in Fig.~\ref{fig mmwave}, clearly showing the discussed advantages. Similar data was obtained for many other mm-wave transitions discussed in this paper. Moreover, when applying an offset electric field, all dipole-allowed transitions between the different Stark-shifted Rydberg M-sublevels are observed at the expected positions. In this case, however, the linewidth of the transitions increases, and is about $5$\,\% of the total Stark shift. This is presumably due to residual electric field inhomogeneities between the capacitor plates.

To obtain improved quantitative results for Rydberg state populations from the mm-wave state transfer, two details can be taken into account. First, mm-wave state transfer is typically performed with relatively high power, ideally equalizing populations among all coupled M-sublevels. However, due to coherences, this is not entirely the case even for very high mm-wave power  where power broadening dominates the spectral linewidth. To eliminate coherences entirely, voltage noise ($\sim100\,$mV$_{\rm rms}$) can be added to the capacitor plates during the mm-wave pulse. This introduces sufficient decoherence to completly equalize populations at sufficiently high mm-wave power, as indicated, among other things, by the amount of state transfer becoming completely power independent for high mm-wave power. Second, in most cases the shown data is based on the change in SFI signal in a time bin optimally containing the ionization peak of the final Rydberg state when switching the mm-wave pulse on/off, relative to the total signal. However, a certain fraction of Rydberg atoms in the final state ionizes outside this time bin, and is thus not counted. This effect is particularly pronounced for final states with principal quantum number $n\gtrsim53$, since here diabatic field ionization at later ionization times starts to dominate. These effects were only understood shortly before taking final measurements shown in sections~\ref{section III.5} and \ref{section III.6}, and are thus only included/taken into account for these results. However, the conclusions drawn from the other results are essentially unaffected.


\section{Results}\label{section III}

\subsection{Signal quality}

\begin{figure}[t]
\centering
\includegraphics[width=.6\textwidth]{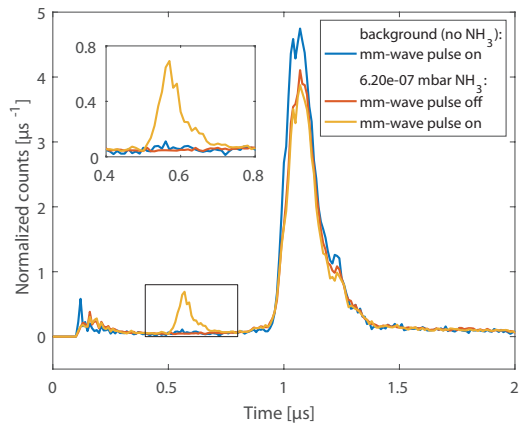}
\caption{SFI signal with and without ammonia, showing F\"orster resonant energy transfer. Rydberg atoms are initially excited to the state $46$P$_{3/2}$. The mm-wave pulse probes the transition $45$D$_{5/2}\leftrightarrow51$P$_{3/2}$.}\label{fig SFIsignal}
\end{figure}

The raw SFI signal showing F\"orster resonant energy transfer between molecules and Rydberg atoms is shown in Fig.~\ref{fig SFIsignal}, based on the experimental sequence and level scheme in Fig.~\ref{fig sequence}. With ammonia present, mm-wave state transfer results in a clear peak corresponding to Rydberg atoms in the $51$P state, whereas either without ammonia or when leaving the mm-wave pulse switched off this peak is absent. The ammonia pressure for the measurements with ammonia corresponds to a density of $1.5\times10^{10}\,$cm$^{-3}$, resulting in a signal to noise ratio of about five for a window of integration between $0.5\,\mu$s and $0.7\,\mu$s. We thus see that F\"orster resonant energy transfer results in robust signals for molecule densities on the order of $10^{10}\,$cm$^{-3}$, and could in fact be used for detection of molecules~\cite{Zeppenfeld17}. This is further discussed in section~\ref{section IV}.

\subsection{Rydberg state maps}

\begin{figure}[t]
\centering
\includegraphics[width=.85\textwidth]{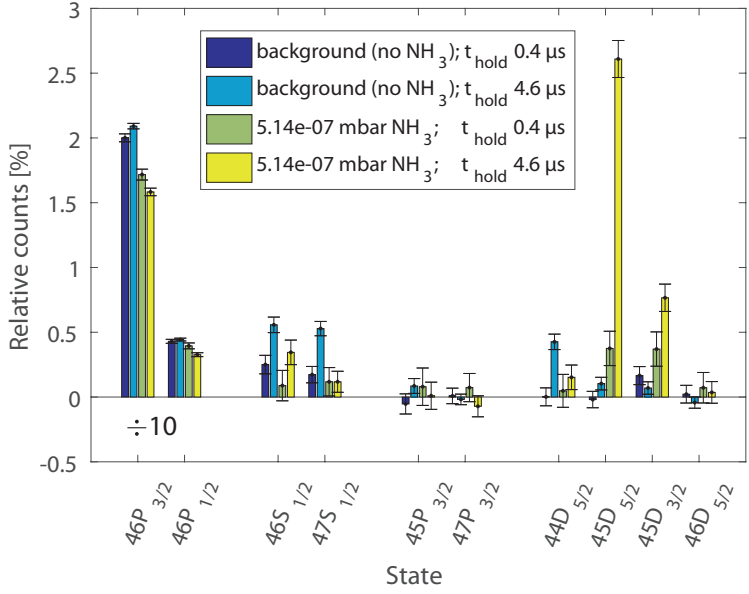}
\caption{Rydberg state populations with and without F\"orster resonant energy transfer. We show the difference in relative SFI signal in the vicinity of the detection peak for the final Rydberg state when switching mm-wave state transfer on/off for different Rydberg states and for four different experimental settings. Specifically, the signal is shown with and without ammonia, in both cases for both a short interaction time $t_{\rm hold}$ and a long interaction time. The data points for the initially excited $46$P states are scaled down by a factor of ten. The mm-wave transitions used for state transfer for the ten states shown are $46P_{3/2}\rightarrow51S_{1/2}$, $46P_{1/2}\rightarrow51S_{1/2}$, $46S_{1/2}\rightarrow50P_{3/2}$, $47S_{1/2}\rightarrow51P_{3/2}$, $45P_{3/2}\rightarrow49S_{1/2}$, $47P_{3/2}\rightarrow52S_{1/2}$, $44D_{5/2}\rightarrow49P_{3/2}$, $45D_{5/2}\rightarrow51P_{3/2}$, $45D_{3/2}\rightarrow51P_{1/2}$, and $46D_{5/2}\rightarrow52P_{3/2}$.}\label{fig statemap}
\end{figure}

While the results in Fig.~\ref{fig SFIsignal} show that population transfer in Rubidium Rydberg atoms due to ammonia takes place as expected for F\"orster resonant energy transfer, they do not show that the transfer is specifically due to F\"orster resonant energy transfer: collisions with ammonia molecules might simply be redistributing the Rydberg atom population to energetically nearby states. We thus investigate the populations in other nearby Rydberg states. Specifically, Fig.~\ref{fig statemap} shows the populations in the $n$S, $n$P, and $n$D states energetically closest to the initially excited $46$P state, both with and without ammonia present and for either a longer interaction time or almost no interaction time.

For all four measurements, the main population is in the $46$P$_{3/2}$ and $46$P$_{1/2}$ states which are initially excited by the UV laser, with almost no population in the other states for a short interaction time. For a long interaction time but without ammonia, population increases in the two $n$S states as well as in the $44D$ state. This is expected from blackbody radiation. The fact that no blackbody transfer is seen for the $45$D states is attributed to the lower transition frequency. Blackbody radiation induced transitions to the $45$D state are observed in the data shown in sections~\ref{section III.5} and \ref{section III.6}, however, due to a longer interaction time and longer data aquisition resulting in improved statistical sensitivity. The fact that no blackbody transfer is seen for the $46$D state is attributed to the much lower transition dipole moment. Less transfer to the states $n$S and $44D$ is observed when ammonia is present, this is attributed to the statistical uncertainty of the measurement.

In contrast to all other state populations, the population in the $45$D states substantially increases for a long interaction time only when ammonia is present. This is expected since, despite the transitions to other $n$S and $n$D states featuring similarly large transition dipole moments, this is the only transition which is resonant to a molecule transition in ammonia. The transition frequencies and transition dipole moments between the $46$P$_{3/2}$ state and all other states in Fig.~\ref{fig statemap} are listed in table~\ref{table}. Since the only thing which strongly distinguishes the $45$D states from all other investigated states is the transition frequency to the $46$P states, this confirms that the observed population transfer is due to F\"orster resonant energy transfer.

\begin{table*}
\centering
\begin{tabular}{l|c|c|l|c|c}
final state:&$f_{\rm trans}$&$d_{\rm trans}$&final state:&$f_{\rm trans}$&$d_{\rm trans}$\\
\hline
$45$P$_{3/2}$&83.6\,GHz&-&$44$D$_{5/2}$&58.3\,GHz&1760\,D\\
$46$P$_{1/2}$&1.07\,GHz&-&$45$D$_{5/2}$&23.6\,GHz&3220\,D\\
$47$P$_{3/2}$&78.0\,GHz&-&$45$D$_{3/2}$&23.5\,GHz&440\,D\\
$46$S$_{1/2}$&40.2\,GHz&2520\,D&$46$D$_{5/2}$&100.1\,GHz&23.6\,D\\
$47$S$_{1/2}$&40.5\,GHz&2460\,D&&\\
\end{tabular}
\caption{Transition frequencies and dipole moments between the $46$P$_{3/2}$ state and all other states in Fig.~\ref{fig statemap}. The dipole moments are calculated for the $M=1/2\leftrightarrow1/2$ transition.}\label{table}
\end{table*}

\subsection{M-sublevel dependence}

\begin{figure}[t]
\centering
\includegraphics[width=.74\textwidth]{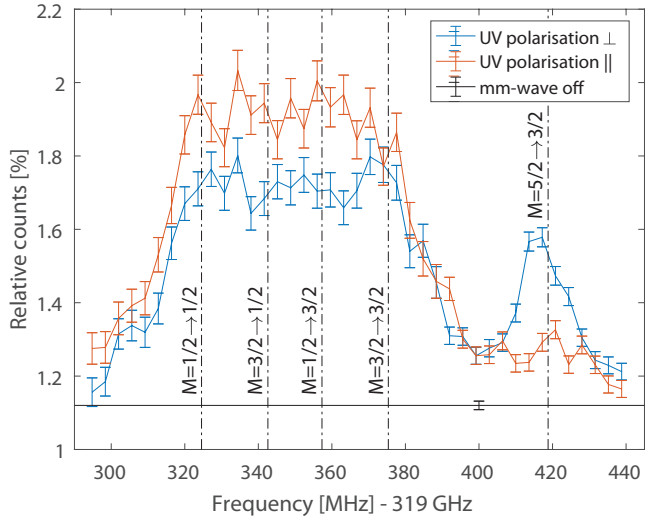}
\caption{M-sublevel dependent F\"orster resonant energy transfer. We show relative SFI signal in the vicinity of the $51$P peak as a function of mm-wave frequency when initially exciting the $46$P state. An offset electric field of approximately $1.03\,$V/cm is applied for this measurement. The vertical lines show the theoretically expected line positions for the labeled M-sublevel components of the $45D_{5/2}\rightarrow51P_{3/2}$ transition for an electric field of $1.03\,$V/cm. The ammonia partial pressure for this measurement is $4.2\times10^{-7}\,$mbar.}\label{fig Msublevels}
\end{figure}

The previous results demonstrated that satisfying the energy resonance condition is important for energy transfer between ammonia and the Rydberg atoms to take place. In this section we present evidence that F\"orster resonant energy transfer also obeys dipole selection rules for the Rydberg M-sublevels. For this purpose, we perform the experimental sequence in Fig.~\ref{fig sequence} with an offset electric field applied. The final population in the $51$P state is measured as a function of mm-wave frequency with reduced mm-wave power to avoid saturation broadening so as to resolve the transitions between individual M-sublevels. Moreover, the experiment was performed twice, once with the UV laser polarization parallel to the applied electric field and once with the UV polarization perpendicular. In the first case we expect Rydberg atoms to be excited with a $\Delta M=0$ selection rule from the $5$S$_{1/2}$ ground state, so that only the $M=\pm1/2$ sublevels are populated. In the second case a $\Delta M=\pm1$ selection rule results in the $M=\pm3/2$ sublevels also being populated. The expected sublevels being populated is verified with mm-wave spectroscopy.

F\"orster resonant energy transfer is expected to follow dipole selection rules $\Delta M=0,\pm1$. As a result we do not expect to see population in $M=\pm5/2$ after interaction with ammonia when only $M=\pm1/2$ sublevels are initially excited by the laser. This is clearly seen in the results shown in Fig.~\ref{fig Msublevels}. When the UV laser polarization is parallel to the applied electric field, mm-wave state transfer from the $45$D$_{5/2}$ state after interaction with ammonia results in a signal when addressing the $M=\pm1/2$ and $M=\pm3/2$ sublevels, but not when addressing the $M=\pm5/2$ sublevels. In contrast, for perpendicular laser polarization, a clear peak is visible when addressing the $M=\pm5/2$ sublevels, indicating that energy transfer to these sublevels takes place as expected.

\subsection{Ammonia pressure dependence}

\begin{figure}[t]
\centering
\includegraphics[width=.7\textwidth]{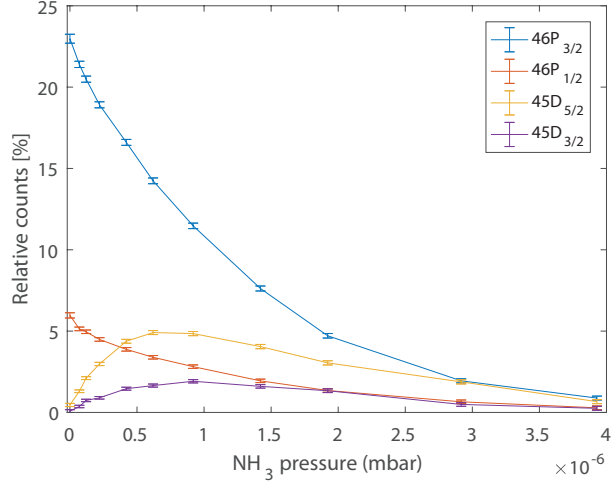}
\caption{Ammonia pressure dependence of F\"orster resonant energy transfer. We show the difference in relative SFI signal in the vicinity of the detection peak for the final Rydberg state when switching mm-wave state transfer on/off. The mm-wave transitions to probe the four Rydberg states are the same as in Fig.~\ref{fig statemap}. The interaction time for this measurement is $8.6\,\mu$s.}\label{fig vspressure}
\end{figure}

We now investigate the ammonia pressure dependence of the observed population transfer. Specifically, we show the amount of mm-wave state transfer from the states $46$P$_{3/2}$, $46$P$_{1/2}$, $45$D$_{5/2}$, and $45$D$_{3/2}$ as a function of ammonia pressure in Fig.~\ref{fig vspressure}.

In the absence of ammonia, population is essentially only observed in the states $46$P$_{3/2}$ and $46$P$_{1/2}$, consistent with Fig.~\ref{fig statemap}. For increasing ammonia pressure up to about $10^{-6}\,$mbar, population in the states $45$D$_{5/2}$ and $45$D$_{3/2}$ increases and population in the states $46$P$_{3/2}$ and $46$P$_{1/2}$ decreases, as expected for F\"orster resonant energy transfer. For a further increase in ammonia pressure, the population in all four states decreases. This is attributed to other collision processes between the molecules and Rydberg atoms~\cite{Matsuzawa83,Dunning83}. In fact, for high ammonia pressure a relatively flat background signal appears during the field-ionization voltage ramp, in contrast to the strongly peaked SFI signal in Figs.~\ref{fig sequence}-\ref{fig SFIsignal}. This indicates that at high ammonia pressure, collisions broadly redistribute the Rydberg atom population across many states. The population in each individual state (other than those shown in Fig.~\ref{fig vspressure}) however remains sufficiently low that detection via mm-wave state transfer is not possible.

The fact that both F\"orster resonant energy transfer and other collision processes become strong for a similar ammonia pressure indicates that the collision cross section for both processes is similar. However, for lower collision energies, the cross section for F\"orster resonant energy transfer is expected to increase dramatically, and even become much larger than the size of the Rydberg electron cloud~\cite{Zeppenfeld17}. In this case, F\"orster resonant energy transfer will dominate.

\subsection{Electric field dependence}\label{section III.5}

\begin{figure}[t]
\centering
\includegraphics[width=1\textwidth]{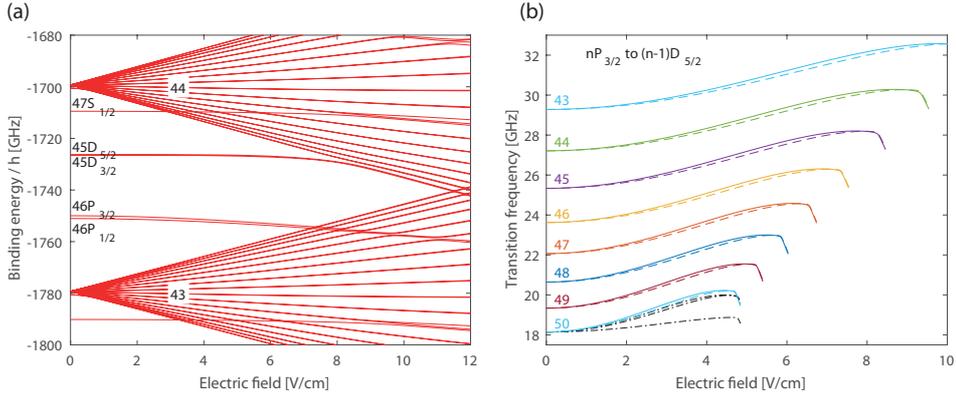}
\caption{Stark shifts for energy levels and transition frequencies of Rubidium Rydberg atoms. (a). Binding energy vs. electric field strength for the $M=1/2$ Rydberg states in the vicinity of the $46$P and $45$D states. Only angular momentum states up to $L=20$ are included in the calculation. (b). Transition frequency for the $n$P$_{3/2}$ to $(n-1)$D$_{5/2}$ transition vs. electric field strength for the values of $n$ as shown. The transitions from $M=1/2$ to $1/2$ are shown in solid, the transitions from $M=1/2$ to $3/2$ are shown dashed. For $n=50$, the transitions from $M=3/2$ in the $50$P state to $M'=1/2, 3/2$ and $5/2$ in the $49$D state are also shown (dash-dot). The Stark shift of the $M=3/2$ to $5/2$ transition deviates substantially from the other transitions.}\label{fig starkmap}
\end{figure}

As a final result, we investigate the dependence of F\"orster resonant energy transfer on an applied electric field. As shown in Fig.~\ref{fig starkmap}a, electric fields on the order of V/cm cause GHz-level shifts for Rydberg states with principal quantum number in the forties. Since F\"orster resonant energy transfer depends on the resonance condition between the Rydberg and molecular transitions, an electric field can be used to shift the Rydberg transition in and out of resonance, thereby changing the energy transfer rate. We note that the applied electric fields cause negligible shifts on the molecule transitions. The applied electric fields also change the transition dipole moments between the Rydberg states. For the transitions and range of electric fields considered, this reduces the transition dipole moment by as much as a factor of two for higher fields~\cite{Patsch18}.

Fig.~\ref{fig starkmap}b shows the shift of the $n$P$_{3/2}$ to $(n-1)$D$_{5/2}$ transition frequency for $n$ in the range from $43$ to $50$. For the $46$P to $45$D transition considered so far, electric fields cause a shift of about $2\,$GHz before the avoided crossings of the $n=43$ manifold of states with the $46$P state cause complications. However, by tuning the electric field and also suitably choosing the principal quantum number, the Rydberg transition frequency can be tuned to any value in the frequency range encompassing the ammonia inversion frequency.

As can be seen in Fig.~\ref{fig starkmap}b, the Rydberg transition frequencies also depend on the Rydberg $M$-sublevels. For the transitions between the $M=1/2$ and $M=3/2$ sublevels, the transition frequencies are relatively similar, and substantially deviate from the transition frequency between the $n$P$_{3/2}$, $M=3/2$ and $(n-1)$D$_{5/2}$, $M'=5/2$ sublevels. In order to obtain a relatively well defined transition frequency for the Rydberg transition when applying an electric field, an offset electric field is applied during Rydberg excitation with the UV laser and the laser polarisation is set parallel to the electric field, thereby mainly populating the $n$P$_{3/2}$, $M=1/2$ Rydberg sublevels. This ideally restricts F\"orster resonant energy transfer to the $n$P$_{3/2}$, $M=1/2$ to $(n-1)$D$_{5/2}$, $M'=1/2$ and $3/2$ transitions.

\begin{figure}[t]
\centering
\includegraphics[width=.6\textwidth]{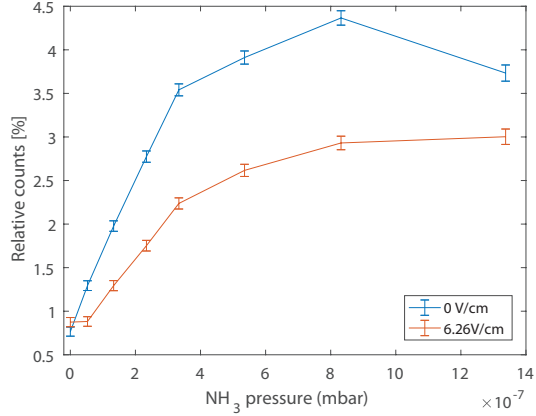}
\caption{Ammonia pressure dependence of F\"orster resonant energy transfer between the $46$P$_{3/2}$ state and the $45$D$_{5/2}$ state for two values of the electric field applied during the interaction time. We show the difference in relative SFI signal in the vicinity of the $51$P state when switching mm-wave state transfer on/off. The interaction time for this measurement is $8.6\,\mu$s, of which the specified electric field is applied for $8.0\,\mu$s.}\label{fig twofieldvsp}
\end{figure}

As a first result, Fig.~\ref{fig twofieldvsp} shows the ammonia pressure dependent population transfer to the $45$D$_{5/2}$ state when applying either a higher electric field or a lower electric field during the interaction time. As can be seen, substantially less population transfer to the $45$D$_{5/2}$ state occurs for higher ammonia pressure for a field of $6.26\,$V/cm than for a field of $0\,$V/cm. This is expected since at zero field the $46$P$_{3/2}$ to $45$D$_{5/2}$ transition is quite close to the center frequency for the ammonia inversion splitting, whereas it shifted away for the higher electric field.

\begin{figure}[t]
\centering
\includegraphics[width=1\textwidth]{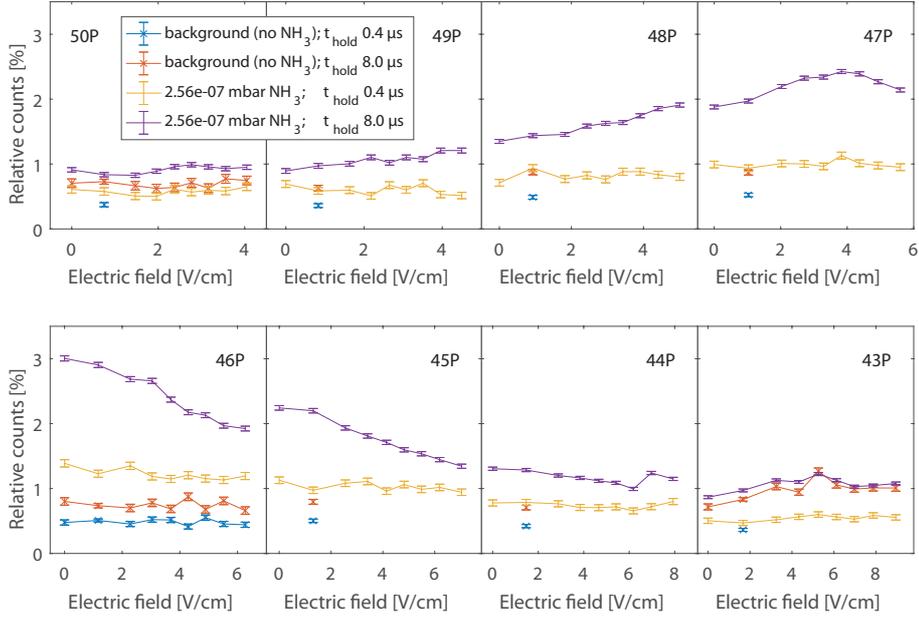}
\caption{Electric field dependence of F\"orster resonant energy transfer between the $n$P$_{3/2}$ state and the $(n-1)$D$_{5/2}$ state. The initially excited Rydberg state $n$P is labeled at the top of each subfigure. We show the difference in relative SFI signal in the vicinity of the detection peak for the final Rydberg state when switching mm-wave state transfer on/off for four different experimental settings as indicated. Note that in contrast to previous measurements, $t_{hold}$ for this measurement only denotes the time during which the specified electric field is applied. An additional $0.6\,\mu$s between the Rydberg excitation and the mm-wave pulse is spent with either the field for Rydberg excitation or the field for the mm-wave pulse applied.}\label{fig rawspectrum}
\end{figure}

To systematically investigate the effect of an electric field on the population transfer, we measure the population transfer both with and without ammonia and for both a short and a long interaction time as a function of the electric field applied during the interaction time for each of the transitions shown in Fig.~\ref{fig starkmap}b. The Rydberg transition frequency is thereby varied from about $18\,$GHz to about $32\,$GHz. In order to suppress events where a Rydberg atom interacts with more than one ammonia molecule, a relatively low ammonia pressure is chosen for this measurement. This will allow us to eliminate the effect of population transfer when the variable electric field is not applied (i.e. directly after Rydberg excitation and during the mm-wave pulse) by taking the difference between the data with a short interaction time and a long interaction time. Since no electric field dependence is expected in the absence of ammonia, data without ammonia is only taken for a single value of the electric field for most of the transitions investigated. The results are shown in Fig.~\ref{fig rawspectrum}.

Without ammonia, the population in the $(n-1)$D$_{5/2}$ state slightly increases for a long interaction time compared to a short interaction time. This is attributed to blackbody radiation. Except when exciting the $43$P state, the data without ammonia shows no statistically significant electric field dependence, as might be expected. The electric field dependence when exciting the $43$P state is discussed below.

When ammonia is present, a clear electric field dependence of the population in the $(n-1)$D$_{5/2}$ state is observed for a long interaction time. Moreover, for increasing electric field strength, the population in the $(n-1)$D$_{5/2}$ state increases for $n=49$ and $48$, peaks for $n=47$, and decreases for $n=46, 45$, and $44$ (disregarding the two points with highest electric field strength for $n=44$). This is precisely what one would expect based on whether the electric field shifts the Rydberg transition closer to or further away from the center frequency of the ammonia inversion splitting. The fact that both an increase and a decrease in population transfer versus electric field strength is observed (depending on $n$) demonstrates that the electric field dependence is predominantly caused by the changing resonance condition for F\"orster resonant energy transfer with ammonia, and not by some other effect such as the change in the transition dipole moment for the Rydberg transition as a function of the electric field. The latter is always similar, independent of $n$.

\subsection{Ammonia inversion spectrum}\label{section III.6}

\begin{figure}[t]
\centering
\includegraphics[width=1\textwidth]{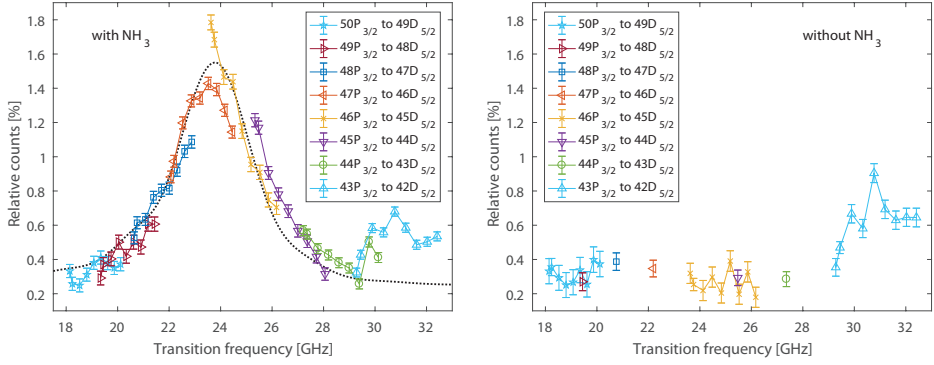}
\caption{Low resolution spectrum of the inversion splitting of ammonia using F\"orster resonant energy transfer. Specifically, we show the difference between the data with long $t_{hold}$ and short $t_{hold}$ in Fig.~\ref{fig rawspectrum} as a function of the Rydberg transition frequency. Since energy transfer to both the $M=1/2$ and $M=3/2$ sublevels of the $(n-1)$D$_{5/2}$ state occurs, the average of the two transition frequencies is used. Since the data with short $t_{hold}$ in Fig.~\ref{fig rawspectrum} shows no statistically significant dependence on the electric field strength, the average value is used for each of the eight datasets to reduce statistical fluctuations. The dotted black line shows a theoretical spectrum of the ammonia inversion splitting as described in the main text.}\label{fig spectrum}
\end{figure}

Combining the data in Fig.~\ref{fig rawspectrum} and plotting as a function of the Rydberg transition frequency allows us to obtain a low resolution spectrum of the ammonia inversion splitting, as shown in Fig.~\ref{fig spectrum}. The data for the different $n$P$_{3/2}$ to $(n-1)$D$_{5/2}$ transitions overlap reasonably well to form a peak centered at the inversion splitting of ammonia. The data for transitions with a given $n$ typically lies below the data with $n-1$, which is consistent with the decreasing transition dipole moment for higher electric fields. Moreover, the data fits quite well to a theoretical spectrum.

The theoretical spectrum is calculated as follows. The frequencies $f_{J,K}$ for the inversion splitting of the different rotational states $J$, $K$ of ammonia are calculated according to $f_{J,K}=f_0-a\,[J(J+1)-K^2]+b\,K^2$ with $f_0=23.786\,$GHz, $a=151.5\,$MHz and $b=59.9\,$MHz according to Ref.~\cite{Townes55}. Each rotational state contributes a gaussian curve centered at $f_{J,K}$ with a full-width-half-maximum of $2\,$GHz to the theoretical spectrum, weighted by the thermal population and the transition dipole moment squared. A vertical offset of $0.25\,$\% is added to account for the signal without ammonia. Vertical scaling and the value for the full-width-half-maximum are chosen by eye to obtain a good fit.

A feature in the data with ammonia which is not accounted for by the theoretical model is the structure observed for a frequency above roughly $29\,$GHz. A very similar feature is observed without ammonia, indicating that the origin is due to some effect other than collisions with ammonia. An explanation is provided by the fact that the capacitor plates in the experiment are separated by about $5\,$mm. As a result, the cutoff frequency for the lowest-order standing-wave mode between the plates will be about $30$\,GHz, and the observed feature is attributed to Purcell enhancement of blackbody induced transitions.

\section{Outlook}\label{section IV}

In summary, we have systematically investigated F\"orster resonant energy transfer between ammonia molecules and rubidium Rydberg atoms. State selective measurements allow key properties of the energy transfer process to be verified, including transition selection rules and dependence on the energy resonance condition. Electric field dependent measurements allow a low-resolution spectrum of the ammonia inversion splitting to be obtained.

Our work suggests many exciting opportunities for future work. As a next step, cold molecules from a velocity filtered source~\cite{Sommer10} might be combined with ultracold atoms in a magneto optical trap. This would allow interactions at roughly two orders-of-magnitude lower collision energies to be investigated, which should result in dramatically increased interaction cross sections~\cite{Zeppenfeld17}. More longterm, ultracold Rydberg atoms might be combined with the more advanced sources of cold and ultracold molecules present in our group~\cite{Zeppenfeld17a}. The combination of buffer-gas cooling with centrifuge deceleration provides motionally and internally cold ensembles of molecules for a wide range of molecules species~\cite{Wu16,Wu17}. Moreover, optoelectrical Sisyphus cooling combined with optical pumping provides state selected ensembles of ultracold formaldehyde~\cite{Prehn16}. Such molecules might be individually loaded into optical tweezers, allowing interactions with individual Rydberg atoms to be investigated. Similar opportunities exist for other sources of cold and ultracold molecules which have been produced in various laboratories worldwide~\cite{Carr09,Lemeshko13}.

A particularly appealing aspect of combining molecules and Rydberg atoms is that observing interactions simultaneously constitutes a detection of the molecules. This is valuable since a low detection efficiency is a key limiting factor for the experiments with cold molecules in our group and elsewhere, and an efficient, generally applicable detection scheme would provide a tremendous boost. The clear signals attributed to interactions with molecules presented in this work shows that Rydberg atoms allow a robust detection of molecules. Here, the sensitivity could be further improved by increasing the interaction time as well as the number of Rydberg atoms produced in a single shot. Interaction time is presently limited by the high velocity of the Rydberg atoms, causing the atoms to leave the detection region. The number of Rydberg atoms produced per shot is presently limited by pile-up in the single-channel channeltron detector. Both these limitations could easily be overcome in future experiments.

In addition to detection of cold and ultracold molecules, interactions with Rydberg atoms might also be used for absolute density measurements in dilute warm gases. Thus, the precise theoretical understanding of Rydberg electron wavefunctions and the ability to accurately approximate the molecule as a rotating point dipole should allow a highly precise calculation of the molecule-Rydberg-atom interaction cross section. Measuring the interaction rate would then allow the absolute molecule density to be determined, which is generally challenging at the densities present in our experiment.

A final outlook is to repeat the electric field dependent measurements presented here at much lower collision energies. This should result in a much narrower resonance condition for energy transfer, resulting in narrow peaks for the individual molecule transitions. In this way, precision spectroscopy of polar molecules using Rydberg atoms instead of photons could be performed.


\ack
Many thanks to Gerhard Rempe for generously supporting this project. Thanks to the rest of the molecule team at MPQ and to Christiane Koch for helpful discussions.\\

\bibliographystyle{unsrt}

\end{document}